# Calculating spin correlations with a quantum computer

Jed Brody, and Gavin Guzman

## ARTICLES YOU MAY BE INTERESTED IN

# Calculating spin correlations with a quantum computer

Jed Brody
*Department of Physics, Emory University, Atlanta, Georgia 30322*

Gavin Guzman
*Department of Computer Science, Emory University, Atlanta, Georgia 30322*

We calculate spin correlation functions using IBM quantum processors, accessed online. We demonstrate the rotational invariance of the singlet state, interesting properties of the triplet states, and surprising features of a state of three entangled qubits. This exercise is ideal for remote learning and generates data with real quantum mechanical systems that are impractical to investigate in the local laboratory. Students learn a wide variety of skills, including calculation of multipartite spin correlation functions, design and analysis of quantum circuits, and remote measurement with real quantum processors. © *2021 American Association of Physics Teachers.*
https://doi.org/10.1119/10.0001967

## I. INTRODUCTION

Quantum computation is a topic of great excitement and promise. Potential applications include cryptography,[1] quantum chemistry,[2] and even responses to the coronavirus pandemic.[3] These factors can motivate students to exert the effort required to learn the abstract mathematics of quantum mechanics. A further benefit of teaching quantum computing is to give students the opportunity to do experiments with real quantum processors via the IBM Quantum Experience.[4] This free platform allows students to design and test real quantum circuits remotely. These experiments illuminate general features of quantum mechanical systems that are otherwise difficult to access, and are ideal for remote learning.

The literature describes a variety of experiments performed with IBM quantum processors. Besides quantum computing algorithms,[5] these experiments include tests of Bell inequalities and Mermin's inequalities for three or more entangled qubits.[6] In this paper, we describe additional experiments of interest to physicists. Although the IBM qubit is not a spin-1/2 particle, both are two-state systems, so the same mathematics applies to both. Any spin direction can be represented by a qubit through use of the Bloch sphere. We use quantum circuits to model the correlation of spins in systems of two and three particles. We demonstrate the rotational invariance of the singlet state, interesting properties of the triplet states, and surprising features of a tripartite state. These experiments aid in the visualization and conceptual understanding of total spin and spin components. Additionally, students gain experience with the theory and experimental use of quantum computers.

## II. SPIN CORRELATIONS

The "computational" basis of a single qubit consists of $|0\rangle$ and $|1\rangle$. On the Bloch sphere (Fig. 1), $|0\rangle$ corresponds with the $+z$ direction, and $|1\rangle$ corresponds with the $-z$ direction. We thus use $|0\rangle$ to represent spin-up in the $z$ direction, $|\uparrow_z\rangle$, and we use $|1\rangle$ to represent spin-down in the $z$ direction, $|\downarrow_z\rangle$. The general Bloch vector

$$\mathbf{n} = \mathbf{i}\sin\theta\cos\phi + \mathbf{j}\sin\theta\sin\phi + \mathbf{k}\cos\theta \quad (1)$$

corresponds with the qubit

$$|\psi\rangle = \cos(\theta/2)|0\rangle + e^{i\phi}\sin(\theta/2)|1\rangle. \quad (2)$$

The appearance of $\theta/2$ in Eq. (2) can be confusing to students and is understood as follows: As $\theta$ goes from 0 ($+z$ direction) to $\pi$ ($-z$ direction), $\cos(\theta/2)$ goes from 1 to 0, and $\sin(\theta/2)$ goes from 0 to 1.

$|\uparrow_z\rangle = |0\rangle$ is represented by the column vector $\begin{pmatrix}1\\0\end{pmatrix}$, and $|\downarrow_z\rangle = |1\rangle$ is represented by $\begin{pmatrix}0\\1\end{pmatrix}$. $\langle 0|$ and $\langle 1|$ are represented by the row vectors (1 0) and (0 1), respectively.

$|\uparrow_z\rangle$ and $|\downarrow_z\rangle$ are eigenvectors of the Pauli spin matrix $\sigma_z = \begin{pmatrix}1 & 0\\0 & -1\end{pmatrix}$, which in quantum computing is usually denoted by Z. Z is the operator associated with measurement of the $z$ component of spin. The measurement of spin in an arbitrary direction $\mathbf{n}$ is associated with the operator $\boldsymbol{\sigma}\cdot\mathbf{n}$, where $\boldsymbol{\sigma} = \sigma_x\mathbf{i} + \sigma_y\mathbf{j} + \sigma_z\mathbf{k}$, with $\sigma_x = X = \begin{pmatrix}0 & 1\\1 & 0\end{pmatrix}$ and $\sigma_y = Y = \begin{pmatrix}0 & -i\\i & 0\end{pmatrix}$. Consequently,

$$\boldsymbol{\sigma}\cdot\mathbf{n} = \begin{pmatrix}n_z & n_x - in_y\\ n_x + in_y & -n_z\end{pmatrix}, \quad (3)$$

where the components of $\mathbf{n}$ are shown in Eq. (1). Students can prove that the eigenvectors of $\boldsymbol{\sigma}\cdot\mathbf{n}$ are $\cos(\theta/2)|0\rangle + e^{i\phi}\sin(\theta/2)|1\rangle$ and $-\sin(\theta/2)|0\rangle + e^{i\phi}\cos(\theta/2)|1\rangle$, with eigenvalues of $+1$ and $-1$, respectively. These eigenvectors correspond with the $\pm\mathbf{n}$ directions on the Bloch sphere.

Any qubit can be written as a superposition of basis states, $a|0\rangle + b|1\rangle$, where $|a|^2 + |b|^2 = 1$ for normalization. When this qubit is measured in the computational basis, 0 is obtained with probability $|a|^2$, and 1 is obtained with probability $|b|^2$. The measurement of a qubit yields a classical bit.

When two qubits are measured in the computational basis, there are four possible outcomes: 00, 01, 10, and 11. The corresponding quantum states form the basis of the two-qubit system and are written $|0\rangle\otimes|0\rangle$, $|0\rangle\otimes|1\rangle$, $|1\rangle\otimes|0\rangle$, and $|1\rangle\otimes|1\rangle$; which may be abbreviated to $|0\rangle|0\rangle$, $|0\rangle|1\rangle$, $|1\rangle|0\rangle$, and $|1\rangle|1\rangle$; or further abbreviated to $|00\rangle$, $|01\rangle$, $|10\rangle$, and $|11\rangle$. $|0\rangle\otimes|0\rangle$ is called a tensor product, a term that can intimidate students. However, when the kets are represented by column vectors, the tensor product is effectively a Kronecker product. The Kronecker product is obtained by straightforward manipulations illustrated below.

If we measure the spin of one particle in the $\boldsymbol{a}$ direction and the spin of a second particle in the $\boldsymbol{b}$ direction, the expectation value of the product of the measurements (neglecting factors of $\hbar/2$) is $\langle\boldsymbol{\sigma}\cdot\boldsymbol{a}\otimes\boldsymbol{\sigma}\cdot\boldsymbol{b}\rangle$. This quantity is the spin correlation function,[7] which appears in Bell's theorem.[8] We wish to calculate the spin correlation function for the triplet states $|00\rangle$, $|11\rangle$, and $\frac{1}{\sqrt{2}}(|01\rangle+|10\rangle)$; and the singlet state $\frac{1}{\sqrt{2}}(|01\rangle-|10\rangle)$. We want to develop a conceptual understanding of the quantitative results. (The significance of the term "triplet states" is that these three states all have the same total spin, though different $z$ components. The singlet state alone has zero total spin.)

The operator $\boldsymbol{\sigma}\cdot\boldsymbol{a}\otimes\boldsymbol{\sigma}\cdot\boldsymbol{b}$ can be written as the Kronecker product of matrices

$$\begin{pmatrix} a_z & a_x-ia_y \\ a_x+ia_y & -a_z \end{pmatrix} \otimes \begin{pmatrix} b_z & b_x-ib_y \\ b_x+ib_y & -b_z \end{pmatrix}.$$

The Kronecker product can be written out as described in numerous texts[9–11] and websites, including Wikipedia. The result in this case is

$$\begin{pmatrix} a_zb_z & a_z(b_x-ib_y) & (a_x-ia_y)b_z & (a_x-ia_y)(b_x-ib_y) \\ a_z(b_x+ib_y) & -a_zb_z & (a_x-ia_y)(b_x+ib_y) & -(a_x-ia_y)b_z \\ (a_x+ia_y)b_z & (a_x+ia_y)(b_x-ib_y) & -a_zb_z & -a_z(b_x-ib_y) \\ (a_x+ia_y)(b_x+ib_y) & -(a_x+ia_y)b_z & -a_z(b_x+ib_y) & a_zb_z \end{pmatrix}.$$

We can now easily determine $\langle\boldsymbol{\sigma}\cdot\boldsymbol{a}\otimes\boldsymbol{\sigma}\cdot\boldsymbol{b}\rangle$ for any state. For $|00\rangle = |0\rangle\otimes|0\rangle = \begin{pmatrix}1\\0\\0\\0\end{pmatrix}$, we have

$$\langle 00|(\boldsymbol{\sigma}\cdot\boldsymbol{a}\otimes\boldsymbol{\sigma}\cdot\boldsymbol{b})|00\rangle = (1\,0\,0\,0)(\boldsymbol{\sigma}\cdot\boldsymbol{a}\otimes\boldsymbol{\sigma}\cdot\boldsymbol{b})\begin{pmatrix}1\\0\\0\\0\end{pmatrix} = a_zb_z$$ 
(4)

by matrix multiplication. The final expression in Eq. (4) can be rewritten as $(\boldsymbol{a}\cdot\mathbf{k})(\boldsymbol{b}\cdot\mathbf{k})$, the product of the projections of the measurement directions on the $+z$ axis. Since $|0\rangle\otimes|0\rangle$ is a product state, the spin correlation function is the product of the expectation values of the single-particle measurements. The expectation value $\langle 11|(\boldsymbol{\sigma}\cdot\boldsymbol{a}\otimes\boldsymbol{\sigma}\cdot\boldsymbol{b})|11\rangle$ is also $a_zb_z$, because the expectation values for the individual particles are $-\boldsymbol{a}\cdot\mathbf{k}$ and $-\boldsymbol{b}\cdot\mathbf{k}$.

For the remaining triplet state $\frac{1}{\sqrt{2}}(|01\rangle+|10\rangle) = \frac{1}{\sqrt{2}}\begin{pmatrix}0\\1\\1\\0\end{pmatrix}$, the expectation value is

$$\frac{1}{2}(0\,1\,1\,0)(\boldsymbol{\sigma}\cdot\boldsymbol{a}\otimes\boldsymbol{\sigma}\cdot\boldsymbol{b})\begin{pmatrix}0\\1\\1\\0\end{pmatrix} = \frac{1}{2}(-2a_zb_z+2a_xb_x+2a_yb_y)$$

$$= \boldsymbol{a}\cdot\boldsymbol{b} - 2(\boldsymbol{a}\cdot\mathbf{k})(\boldsymbol{b}\cdot\mathbf{k}).$$ 
(5)

Since this simplifies to $\boldsymbol{a}\cdot\boldsymbol{b}$ when both particles are measured in the $x$-$y$ plane, we see that the spins are correlated along any direction in the $x$-$y$ plane. However, the spins are anticorrelated along the $z$ axis, which is necessary for a superposition of $|01\rangle$ and $|10\rangle$.

Finally, the singlet state $\frac{1}{\sqrt{2}}(|01\rangle-|10\rangle) = \frac{1}{\sqrt{2}}\begin{pmatrix}0\\1\\-1\\0\end{pmatrix}$ has the expectation value

$$\frac{1}{2}(0\,1\,-1\,0)(\boldsymbol{\sigma}\cdot\boldsymbol{a}\otimes\boldsymbol{\sigma}\cdot\boldsymbol{b})\begin{pmatrix}0\\1\\-1\\0\end{pmatrix}$$

$$= \frac{1}{2}(-2a_zb_z-2a_xb_x-2a_yb_y) = -\boldsymbol{a}\cdot\boldsymbol{b}.$$ 
(6)

The spins are anticorrelated along any direction, because the total spin of the singlet state is 0. The singlet state is famously analyzed in Bell's theorem and is an entangled state, because it cannot be written as a product of the states of the individual particles. The triplet state $1/\sqrt{2}(|01\rangle+|10\rangle)$ is also an entangled state.

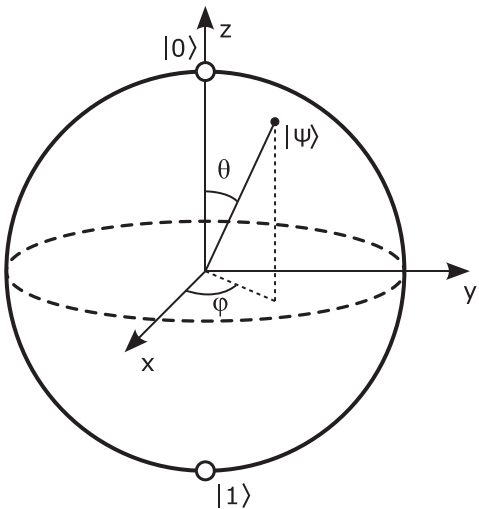

Fig. 1. The Bloch sphere, reproduced exactly from https://commons.wikimedia.org/wiki/File:Bloch_sphere.svg.

We will focus on the x-z and x-y planes in the Bloch sphere. In the x-z plane ($\phi = 0$ so that $\boldsymbol{n} = \boldsymbol{i}\sin\theta + \boldsymbol{k}\cos\theta$), we define

$$|\uparrow_\theta\rangle \equiv \cos(\theta/2)|0\rangle + \sin(\theta/2)|1\rangle \tag{7}$$

and

$$|\downarrow_\theta\rangle \equiv -\sin(\theta/2)|0\rangle + \cos(\theta/2)|1\rangle, \tag{8}$$

the eigenvectors of

$$\sigma_\theta \equiv X\sin\theta + Z\cos\theta = \begin{pmatrix} \cos\theta & \sin\theta \\ \sin\theta & -\cos\theta \end{pmatrix}. \tag{9}$$

In the x-y plane ($\theta = \pi/2$ so that $\boldsymbol{n} = \boldsymbol{i}\cos\phi + \boldsymbol{j}\sin\phi$), we define

$$|\uparrow_\phi\rangle \equiv \frac{1}{\sqrt{2}}\left(|0\rangle + e^{i\phi}|1\rangle\right) \tag{10}$$

and

$$|\downarrow_\phi\rangle \equiv \frac{1}{\sqrt{2}}\left(|0\rangle - e^{i\phi}|1\rangle\right), \tag{11}$$

the eigenvectors of

$$\sigma_\phi \equiv X\cos\phi + Y\sin\phi = \begin{pmatrix} 0 & e^{-i\phi} \\ e^{i\phi} & 0 \end{pmatrix}. \tag{12}$$

We now apply Eqs. (4)–(6) to determine $\langle Z\otimes\sigma_\theta\rangle$, $\langle\sigma_\theta\otimes\sigma_\theta\rangle$, $\langle X\otimes\sigma_\phi\rangle$, and $\langle\sigma_\phi\otimes\sigma_\phi\rangle$. These are the spin correlation functions that we will determine experimentally with a quantum computer.

$\langle Z\otimes\sigma_\theta\rangle$ is $\cos\theta$ for the product states and $-\cos\theta$ for the entangled states. $\langle\sigma_\theta\otimes\sigma_\theta\rangle$ is $\cos^2\theta$ for the product states, $1 - 2\cos^2\theta$ for the other triplet state, and $-1$ for the singlet state. $\langle X\otimes\sigma_\phi\rangle$ is 0 for the product states, $\cos\phi$ for the other triplet state, and $-\cos\phi$ for the singlet state. Finally, $\langle\sigma_\phi\otimes\sigma_\phi\rangle$ is 0 for the product states, 1 for the other triplet state, and $-1$ for the singlet state.

This is just a sampling of spin correlations that students can calculate. Measurements are not limited to the x-z and x-y planes, and states are not limited to eigenstates of the total spin operator. Students could investigate $\frac{1}{\sqrt{2}}(|00\rangle+|11\rangle)$, for example, as part of an open-ended exploration.

Tripartite states offer even more opportunities for mathematical and experimental investigation. We give just one interesting example. We examine the W state,[12] $|W\rangle = \frac{1}{\sqrt{3}}(|001\rangle + |010\rangle + |100\rangle)$, when each particle is measured along the same direction in the x-z plane. We need to calculate $\langle W|(\sigma_\theta\otimes\sigma_\theta\otimes\sigma_\theta)|W\rangle$. It is possible to write out the Kronecker product as an $8 \times 8$ matrix. Alternatively, it is possible to evaluate each of nine terms, one of which is

$$((1\,0)\otimes(1\,0)\otimes(0\,1))\left(\begin{pmatrix}\cos\theta & \sin\theta \\ \sin\theta & -\cos\theta\end{pmatrix}\right.$$
$$\left.\otimes\begin{pmatrix}\cos\theta & \sin\theta \\ \sin\theta & -\cos\theta\end{pmatrix}\otimes\begin{pmatrix}\cos\theta & \sin\theta \\ \sin\theta & -\cos\theta\end{pmatrix}\right)$$
$$\times\left(\begin{pmatrix}1\\0\end{pmatrix}\otimes\begin{pmatrix}1\\0\end{pmatrix}\otimes\begin{pmatrix}0\\1\end{pmatrix}\right)$$
$$= ((1\,0)\otimes(1\,0)\otimes(0\,1))\left(\begin{pmatrix}\cos\theta\\\sin\theta\end{pmatrix}\otimes\begin{pmatrix}\cos\theta\\\sin\theta\end{pmatrix}\right.$$
$$\left.\otimes\begin{pmatrix}\sin\theta\\-\cos\theta\end{pmatrix}\right) = -\cos^3\theta.$$

Evaluating all terms we obtain $\langle W|(\sigma_\theta\otimes\sigma_\theta\otimes\sigma_\theta)|W\rangle = \cos\theta(2 - 3\cos^2\theta)$. There is no obvious conceptual explanation for this spin correlation function, but it is an interesting function to derive and compare with experiment. Students can work to discover other interesting spin correlation functions in systems of three or more particles.

Students may find that these analytical calculations are more tedious than the use of a quantum computer to obtain experimental results. Indeed, this is part of the reason quantum computers have the potential to model quantum systems more efficiently than classical computers do.

## III. QUANTUM GATES, INITIAL STATES, AND BLOCH VECTOR ROTATIONS

Accessible introductions to quantum computing are found elsewhere.[5,9–11,13,14] Here, we list only the operators, called quantum gates, which we use in our experiments. Besides the Pauli spin matrices, we need just a handful of other quantum gates. The Hadamard gate,

$$H = \frac{1}{\sqrt{2}}\begin{pmatrix}1 & 1 \\ 1 & -1\end{pmatrix}, \tag{13}$$

transforms computational basis states into superpositions. Gates that effect rotations around the y and z axes of the Bloch sphere are

$$R_y(\theta) = \begin{pmatrix}\cos\dfrac{\theta}{2} & -\sin\dfrac{\theta}{2} \\ \sin\dfrac{\theta}{2} & \cos\dfrac{\theta}{2}\end{pmatrix} \tag{14}$$

and

$$R_z(\phi) = \begin{pmatrix}e^{-i\phi/2} & 0 \\ 0 & e^{i\phi/2}\end{pmatrix}, \tag{15}$$

respectively. We will also use a two-qubit gate, controlled NOT,

$$\text{CNOT} = \begin{pmatrix}1 & 0 & 0 & 0 \\ 0 & 1 & 0 & 0 \\ 0 & 0 & 0 & 1 \\ 0 & 0 & 1 & 0\end{pmatrix}. \tag{16}$$

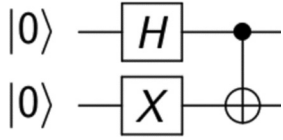

Fig. 2. Circuit to create $\frac{1}{\sqrt{2}}(|01\rangle + |10\rangle)$.

CNOT acts as a NOT on the second qubit (the target) only when the first qubit (the control) is in the $|1\rangle$ state. Finally, we will need the three-qubit Toffoli gate, which acts as a NOT on the third qubit only when the first two are in the $|1\rangle$ state.

Qubits are initialized to $|0\rangle$ by default in the IBM Quantum Experience. We need to create the states analyzed in Sec. II. To create $|1\rangle$, we simply use the X gate since $X|0\rangle = |1\rangle$. To create $\frac{1}{\sqrt{2}}(|01\rangle + |10\rangle)$, we use the circuit shown in Fig. 2. Quantum circuits are read from left to right. The H gate puts the upper qubit in the state $\frac{1}{\sqrt{2}}(|0\rangle + |1\rangle)$, and the X gate puts the lower qubit in the state $|1\rangle$. The final symbol is the controlled NOT, with the upper qubit as the control and the lower qubit as the target. Just before the controlled NOT, the state of the qubits is $H|0\rangle \otimes X|0\rangle = \frac{1}{\sqrt{2}}(|0\rangle + |1\rangle) \otimes |1\rangle = \frac{1}{\sqrt{2}}(|01\rangle + |11\rangle)$, where we write the upper qubit first. (This is the convention followed in most textbooks. However, the IBM Quantum Experience follows the opposite convention and writes the lower qubit first.) The controlled NOT transforms $\frac{1}{\sqrt{2}}(|01\rangle + |11\rangle)$ into $\frac{1}{\sqrt{2}}(|01\rangle + |10\rangle)$. The circuit to create the singlet state appears to the left of the broken line in Fig. 3, as students can confirm.

Figure 4 shows the circuit to create the W state, $\frac{1}{\sqrt{3}}(|001\rangle + |010\rangle + |100\rangle)$. It would be challenging to design this circuit[15] from scratch, but students can learn how it works. After the $R_y$ gate, defined in Eq. (14), the state is $\left(\frac{1}{\sqrt{3}}|0\rangle + \sqrt{\frac{2}{3}}|1\rangle\right) \otimes |0\rangle \otimes |0\rangle = \frac{1}{\sqrt{3}}|000\rangle + \sqrt{\frac{2}{3}}|100\rangle$. The H gate is a controlled H gate, such that H is applied to the second qubit only if the first qubit is $|1\rangle$. After this gate, the state is $\frac{1}{\sqrt{3}}(|000\rangle + |100\rangle + |110\rangle)$. The next symbol is a Toffoli gate, which transforms the state to $\frac{1}{\sqrt{3}}(|000\rangle + |100\rangle + |111\rangle)$, after which the X gates make it $\frac{1}{\sqrt{3}}(|110\rangle + |010\rangle + |001\rangle)$. Finally, the controlled NOT gate produces the desired state.

The next question is how to experimentally determine the expectation values calculated in Sec. II. This is an important question, and students should consider it carefully. The IBM Quantum Experience allows measurements in the computational basis only. A measurement of two qubits yields one of four possible results: 00, 01, 10, or 11. Each experiment is repeated many times, and the experimental data are the numbers of occurrences of each outcome. The probabilities of obtaining the possible outcomes are then estimated as the experimental fractions of occurrence.

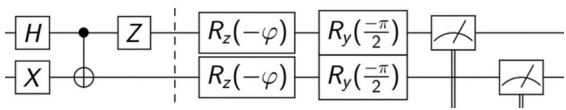

Fig. 3. Circuit to create the singlet state, $\frac{1}{\sqrt{2}}(|01\rangle - |10\rangle)$, and determine $\langle \sigma_\phi \otimes \sigma_\phi \rangle$. The initial state is $|00\rangle$ by default. The broken line separates the creation of the state from the measurement process. The rotation operators effect a change of basis, as shown in Eq. (20). The symbols on the far right represent measurements in the computational basis.

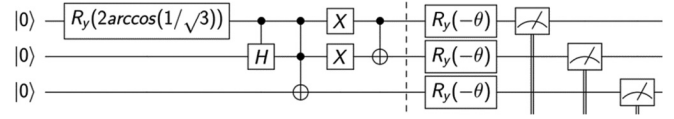

Fig. 4. Circuit to create $|W\rangle = \frac{1}{\sqrt{3}}(|001\rangle + |010\rangle + |100\rangle)$ and determine $\langle W|(\sigma_\theta \otimes \sigma_\theta \otimes \sigma_\theta)|W\rangle$.

Suppose $|\psi\rangle$ is a two-qubit state, and we wish to determine $\langle\psi|U_1 \otimes U_2|\psi\rangle$, where $U_1$ and $U_2$ are unitary operators acting on the first and second qubit, respectively. We define the orthogonal eigenvectors of these operators through $U_i|\uparrow_i\rangle = |\uparrow_i\rangle$ and $U_i|\downarrow_i\rangle = -|\downarrow_i\rangle$, where $i$ is 1 or 2, and we see that the eigenvalues are $\pm 1$. Next we write $|\psi\rangle$ in terms of the eigenvectors of $U_1 \otimes U_2$

$$|\psi\rangle = p|\uparrow_1\rangle \otimes |\uparrow_2\rangle + q|\uparrow_1\rangle \otimes |\downarrow_2\rangle + r|\downarrow_1\rangle \otimes |\uparrow_2\rangle + s|\downarrow_1\rangle \otimes |\downarrow_2\rangle, \quad (17)$$

where $p$, $q$, $r$, and $s$ are the amplitudes. Since the eigenvalues of $|\uparrow_1\rangle \otimes |\uparrow_2\rangle$ and $|\downarrow_1\rangle \otimes |\downarrow_2\rangle$ are 1, and the eigenvalues of $|\uparrow_1\rangle \otimes |\downarrow_2\rangle$ and $|\downarrow_1\rangle \otimes |\uparrow_2\rangle$ are $-1$, the expectation value must be the probability of obtaining 1 minus the probability of obtaining $-1$

$$\langle\psi|U_1 \otimes U_2|\psi\rangle = |p|^2 + |s|^2 - |q|^2 - |r|^2. \quad (18)$$

When the inner product with $|\uparrow_1\rangle \otimes |\uparrow_2\rangle$ is formed on of both sides of Eq. (17), we find $p = (\langle\uparrow_1|\otimes\langle\uparrow_2|)|\psi\rangle$. Similar expressions are obtained for $q$, $r$, and $s$, so that Eq. (18) becomes

$$\langle\psi|U_1 \otimes U_2|\psi\rangle = |(\langle\uparrow_1|\otimes\langle\uparrow_2|)|\psi\rangle|^2 + |(\langle\downarrow_1|\otimes\langle\downarrow_2|)|\psi\rangle|^2 - |(\langle\uparrow_1|\otimes\langle\downarrow_2|)|\psi\rangle|^2 - |(\langle\downarrow_1|\otimes\langle\uparrow_2|)|\psi\rangle|^2. \quad (19)$$

Next, we define the unitary operator $R_i$ that maps the eigenvectors of $U_i$ onto the computational basis: $R_i|\uparrow_i\rangle = |0\rangle$ and $R_i|\downarrow_i\rangle = |1\rangle$, so that $\langle\uparrow_i|R_i^\dagger = \langle 0|$ and $\langle\downarrow_i|R_i^\dagger = \langle 1|$. Then since $R_i^\dagger R_i = 1$, $|(\langle\uparrow_1|\otimes\langle\uparrow_2|)|\psi\rangle|^2 = |(\langle\uparrow_1|\otimes\langle\uparrow_2|)R_1^\dagger R_1 \otimes R_2^\dagger R_2|\psi\rangle|^2 = |(\langle\uparrow_1|R_1^\dagger \otimes \langle\uparrow_2|R_2^\dagger) R_1 \otimes R_2|\psi\rangle|^2 = |(\langle 0|\otimes\langle 0|)R_1 \otimes R_2|\psi\rangle|^2 = |\langle 00|R_1 \otimes R_2|\psi\rangle|^2$. Transforming each term in Eq. (19) this way

$$\langle\psi|U_1 \otimes U_2|\psi\rangle = |\langle 00|R_1 \otimes R_2|\psi\rangle|^2 + |\langle 11|R_1 \otimes R_2|\psi\rangle|^2 - |\langle 01|R_1 \otimes R_2|\psi\rangle|^2 - |\langle 10|R_1 \otimes R_2|\psi\rangle|^2. \quad (20)$$

These four terms are the probabilities of obtaining 00, 11, 01, and 10, when the state $R_1 \otimes R_2|\psi\rangle$ is measured in the computational basis. So the experimental determination of $\langle\psi|U_1 \otimes U_2|\psi\rangle$ requires the application of $R_1 \otimes R_2$ to $|\psi\rangle$, followed by measurement in the computational basis. In effect, measuring $R_1 \otimes R_2|\psi\rangle$ in the computational basis is the same as measuring $|\psi\rangle$ in the basis of the eigenvectors of $U_1 \otimes U_2$.

$R_i$, the operator that maps eigenvectors of $U_i$ onto the computational basis, can be determined from simple geometric considerations. Suppose we wish to measure along the $x$ axis of the Bloch sphere. If we were able to make this

38      Am. J. Phys., Vol. 89, No. 1, January 2021      J. Brody and G. Guzman      38

measurement directly, the result would be one of the eigenvectors of X, corresponding with the Bloch vectors $\pm\mathbf{i}$. We need an operator that maps the eigenvectors of X onto the eigenvectors of Z, effectively mapping $\pm\mathbf{i}$ onto $\pm\mathbf{k}$. $R_y(-\pi/2)$ performs precisely this operation. More generally, $R_y(-\theta)$ maps $|\uparrow_\theta\rangle$ to $|0\rangle$ and $|\downarrow_\theta\rangle$ to $|1\rangle$. Skeptical students are encouraged to prove these facts by use of Eqs. (7), (8), and (14). The complete circuit to determine $\langle W|(\sigma_\theta \otimes \sigma_\theta \otimes \sigma_\theta)|W\rangle$ is shown in Fig. 4.

To measure in the *x-y* plane of the Bloch sphere, we need to map $|\uparrow_\phi\rangle$ to $|0\rangle$ and $|\downarrow_\phi\rangle$ to $|1\rangle$. We recognize that we first need to apply $R_z(-\phi)$ to rotate the Bloch vector onto the *x* axis, and then we apply $R_y(-\pi/2)$ to rotate the Bloch vector onto the *z* axis. The complete circuit to determine $\langle\sigma_\phi\otimes\sigma_\phi\rangle$ for the singlet state is shown in Fig. 3.

## IV. EXPERIMENT

The IBM Quantum Experience provides two options for creating and running quantum circuits: the Circuit Composer, which is a graphical interface, and Qiskit, which is a Python package that can be used either locally or on the IBM website. The Circuit Composer is easy to learn and immediately displays circuits like those shown in Figs. 2–4. All of our experiments can be performed in the Circuit Composer, though each value of $\theta$ or $\phi$ must be entered manually. This tedium is eliminated by using for-loops in Qiskit. The Qiskit statements that add quantum gates are simple, and the rest of the code can be provided to students as a template.[16]

We created quantum circuits to experimentally determine $\langle Z\otimes\sigma_\theta\rangle$, $\langle\sigma_\theta\otimes\sigma_\theta\rangle$, $\langle X\otimes\sigma_\phi\rangle$, and $\langle\sigma_\phi\otimes\sigma_\phi\rangle$, for the states $|00\rangle$, $|11\rangle$, $\frac{1}{\sqrt{2}}(|01\rangle+|10\rangle)$, and $\frac{1}{\sqrt{2}}(|01\rangle-|10\rangle)$. Each circuit was run 1024 times on the quantum processor called "ibmq_ourense," and the expectation values were estimated, as described in Sec. III, as the fraction of 00 and 11 results, minus the fraction of 01 and 10 results. When measuring in the *x-z* plane of the Bloch sphere, we varied $\theta$ from 0 to $\pi$ in intervals of $\pi/16$. When measuring in the *x-y* plane of the Bloch sphere, we varied $\phi$ from 0 to $2\pi$ in intervals of $\pi/8$. Additionally, we created a circuit to estimate $\langle W|(\sigma_\theta \otimes \sigma_\theta \otimes \sigma_\theta)|W\rangle$ as the fraction of 000, 011, 101, and 110 results, minus the fraction of 001, 010, 100, and 111 results. In this case, we varied $\theta$ from 0 to $\pi$ in intervals of $\pi/18$.

Results for two-particle spin correlation functions are shown in Figs. 5–8. In all cases, the experimental data generally follow the expected trends. However, the experimental expectation values are never get quite as high as 1 or quite as low as −1. Random error has this tendency to reduce the magnitude of the expectation value.

Another systematic error appears in Fig. 5. The experimental results are mostly higher than expected for $|11\rangle$, but lower than expected for $|00\rangle$. This result is consistent with an $R_y(\theta)$ operation that produces larger than intended amplitudes of $|1\rangle$, but we do not know if this is the cause of the error.

Figure 5 shows that $|00\rangle$ and $|11\rangle$ results are correlated when both qubits are measured in the same direction along the *z* axis of the Bloch sphere, but they are anticorrelated when measured in opposite directions. The correlation function decreases smoothly as the angle between measurement directions increases from 0 to $\pi$. In contrast, for the superpositions of $|01\rangle$ and $|10\rangle$, the correlation function increases as expected.

Figure 6 shows again that $|00\rangle$ and $|11\rangle$ results are correlated when the qubits are measured in the same direction along the *z* axis ($\theta=0$ and $\theta=\pi$), and also that the results are uncorrelated when the qubits are measured along the *x* axis ($\theta=\pi/2$). The singlet results are always anticorrelated when both qubits are measured in the same direction in the Bloch sphere. In contrast, results for $\frac{1}{\sqrt{2}}(|01\rangle+|10\rangle)$ are correlated when both qubits are measured along the *x* axis.

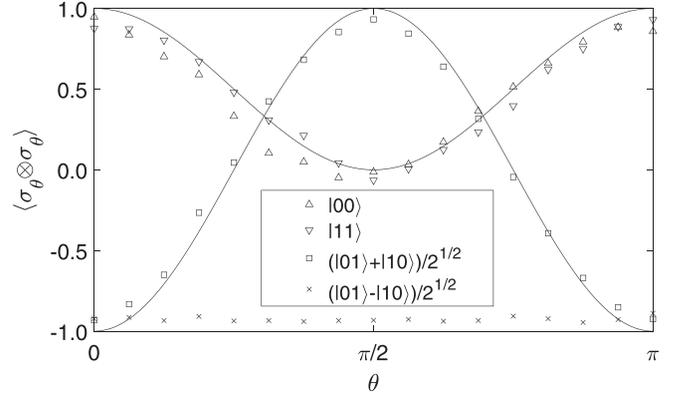

Fig. 6. Spin correlation functions with both qubits measured along the same line in the *x-z* plane of the Bloch sphere. The solid lines are theory, and the theoretical expectation value for the singlet state is a constant −1.

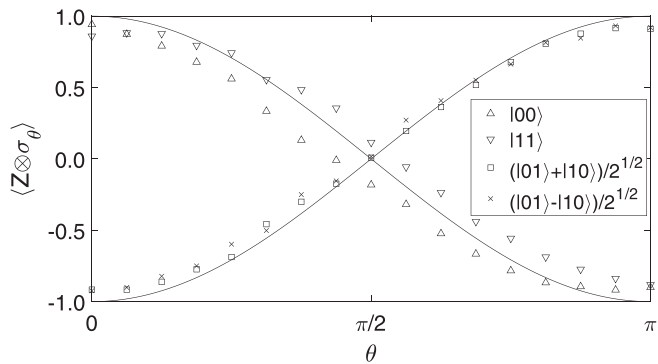

Fig. 5. Spin correlation functions with one qubit measured in the computational basis and the other measured in the *x-z* plane of the Bloch sphere. The solid lines are theory.

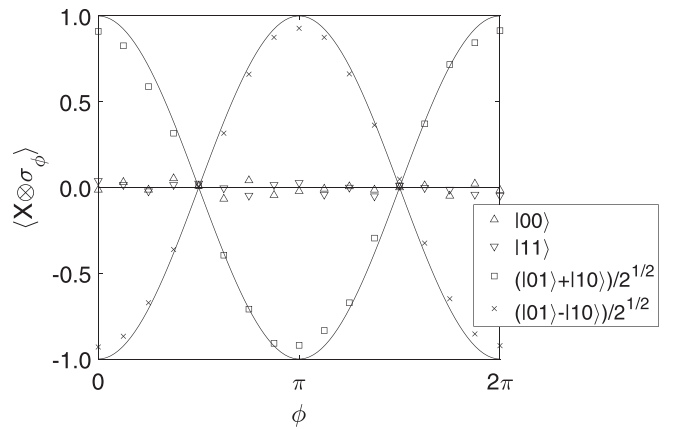

Fig. 7. Spin correlation functions with one qubit measured along the *x* direction and the other measured in the *x-y* plane of the Bloch sphere. The solid lines are theory.

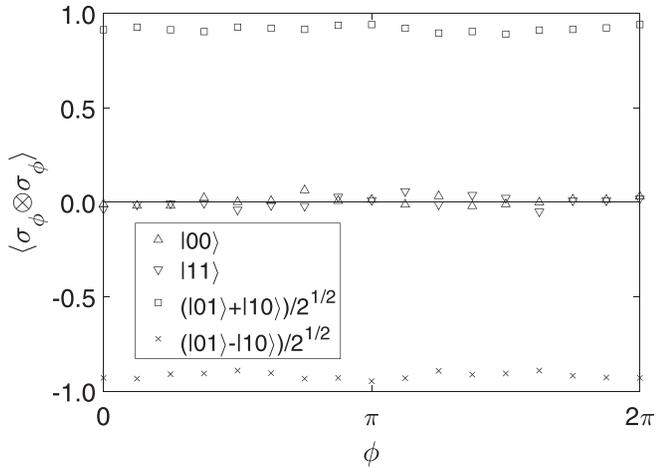

Fig. 8. Spin correlation functions with both qubits measured along the same line in the *x-y* plane of the Bloch sphere.

Figures 7 and 8 show that $|00\rangle$ and $|11\rangle$ results are uncorrelated, when both qubits are measured in the *x-y* plane of the Bloch sphere. In the *x-y* plane, the smaller the least angle between measurement directions, the greater the correlation of the results for $\frac{1}{\sqrt{2}}(|01\rangle + |10\rangle)$. In any plane, the smaller the least angle between measurement directions, the greater the anticorrelation for the singlet.

Figure 9 shows results for the three-qubit W state. There is not an obvious conceptual explanation for the relative extrema near $\pi/3$ and $2\pi/3$, but this surprising feature is confirmed by experiment. Instructors may wish to have students obtain experimental results first and then wonder, "How do I derive an equation for this?"

We found that another quantum processor, ibmq_london, gave inferior results in this experiment, although the calibration data for ibmq_london and ibmq_ourense were similar.[18] In any case, the calibration data are updated almost daily,[19] and we do not know whether ibmq_ourense always performs better than ibmq_london.

Quantum-computing error, along with its correction, is a deep subject of active research. Readout errors, gate errors, and decoherence all affect experimental results.[20] Measured results can be compared with a noise model[21] if desired. A further complication is the "transpiling" of quantum circuit diagrams into gates that are available in the hardware.[22] For example, the Toffoli gate is constructed from six or more CNOT gates and additional single-qubit gates.[23]

## V. CONCLUSIONS

Earlier work introduced laboratory experiments into the quantum mechanics curriculum.[17] We now have an opportunity to incorporate remote quantum-computing experiments as well. These experiments illustrate universal features of quantum mechanical systems, including entanglement, measurement probabilities, and unitary operations for basis transformations. Students are able to probe states that are impractical to investigate in the local laboratory. Students learn a wide variety of skills, including calculation of spin correlation functions, design and analysis of quantum circuits, and remote measurement with real quantum processors.

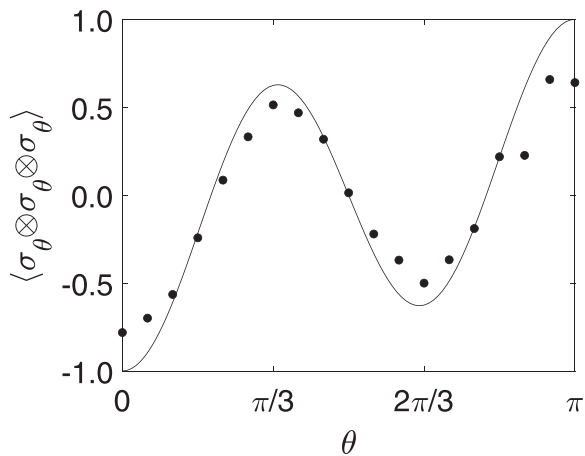

Fig. 9. Spin correlation function for the W state with each qubit measured along the same line of the *x-z* plane of the Bloch sphere. The solid line is theory.